\documentclass[aps,twocolumn,prl]{revtex4}
\usepackage{graphicx}% Include figure files
\usepackage{dcolumn}% Align table columns on decimal point
\usepackage{bm}% bold math
\usepackage{amsmath}
\usepackage{color}

\begin{document}
\preprint{APS/123-QED}

\title{Uniaxial stress tuning of geometrical frustration in a Kondo lattice}

\author{R. K\"{u}chler$^{1,2}$}
\author{C. Stingl$^2$}
\author{Y. Tokiwa$^2$}
\author{M. S. Kim$^3$}
\author{T. Takabatake$^3$}
\author{P. Gegenwart$^2$}

\affiliation{$^1$Max Planck Institute for Chemical Physics of Solids, 01187 Dresden, Germany}
\affiliation{$^2$Experimental Physics VI, Center for Electronic Correlations and Magnetism, University of Augsburg, 86159 Augsburg, Germany}
\affiliation{$^3$Department of Quantum Matter, ADSM, Hiroshima University, Higashi-Hiroshima, 739-8530, Japan}

\date{\today}% It is always \today, today,
% but any date may be explicitly specified
\pacs{}% PACS, the Physics and Astronomy

\begin{abstract}
Hexagonal CeRhSn with paramagnetic $4f$ moments on a distorted Kagome lattice displays zero-field quantum critical behavior related to geometrical frustration. We report high-resolution thermal expansion and magnetostriction measurements under multiextreme conditions such as uniaxial stress up to 200~MPa, temperatures down to 0.1~K and magnetic fields up to 10~T. Under uniaxial stress along the $a$-direction, quantum criticality disappears and a complex magnetic phase diagram arises with a sequence of phases below 1.2~K and fields between 0~and 3~T ($\parallel a$). Since the Kondo coupling increases with stress, which alone would stabilize paramagnetic behavior in CeRhSn, the observed order arises from the release of geometrical frustration by in-plane stress.
%our data directly prove that geometrical frustration acts as second control parameter for tuning quantum criticality in heavy-fermion metals as proposed previously in the ``global phase diagram''.
\end{abstract}

\maketitle

A quantum critical point (QCP) denotes a second-order phase transformation at zero temperature, driven by a non-thermal control parameter. QCPs have been discussed for various areas of physics in recent years~\cite{sachdev:qcp-book}. Kondo lattices, consisting of localized magnetic moments and conduction electrons, show particularly fascinating phenomena near the QCP, due to the interaction of magnetic and electronic degrees of freedom~\cite{LohneysenHilbertV:Ferimq}. Particularly striking is the observation of non-Fermi liquid (NFL) effects, which cannot be described by the standard itinerant theory of quasiparticles that are scattered by critical magnetic fluctuations. It is well experimentally established, that Kondo lattice systems can be tuned across the QCP by effectively changing the coupling strength ($J$) between local $f$- and conduction electrons through pressure, magnetic field or chemical substitution~\cite{gegenwart-review}. More recently, it has been proposed that even without changing $J$ the introduction of strong geometrical frustration suppresses magnetic order leading to a novel metallic spin liquid like ground state~\cite{Si-PhysicaB06,Vojta-prb08,Coleman-JLTP10}. The ``global phase diagram'' for Kondo lattices distinguishes four different states AF$_{\rm L}$, AF$_{\rm S}$, PM$_{\rm L}$ and PM$_{\rm S}$ (where AF and PM denote antiferromagnetic and paramagnetic behavior while the subscripts L and S refer to large and small Fermi surface volume, i.e., itinerant or localized $f$-electrons, respectively) and two complementary tuning parameters $J$ and $Q$. Here $Q$ denotes the strength of quantum fluctuations induced by geometrical frustration, although it is difficult to quantify this parameter. Up to now there exists no example for the tuning of a Kondo lattice through its QCP by a variation of $Q$. The reason for this is that geometrical frustration cannot be varied systematically using established control parameters. Below, we report important progress in this direction.

%However, since geometric frustration sensitively depends on slight distortions, external stress applied along special crystallographic directions should be a possibility to substantially weaken frustration.
The effect of geometrical frustration in Kondo lattices has rarely been observed experimentally. Highlights include the observation of partial magnetic order in \mbox{CePdAl} \cite{Doenni-JPhys96}, quantum criticality in pyrochlore Pr$_2$Ir$_2$O$_7$~\cite{Tokiwa-NM14} and spinon-type excitations in the Shastry-Sutherland lattice system Yb$_2$Pt$_2$Pb~\cite{Aronson2016}. In hexagonal CePdAl the $4f$ moments are located on equilateral corner-sharing triangles in the $ab$ plane, forming a distorted kagome network. At low temperatures, two-third of Ce moments order, while one-third remain disordered due to geometrical frustration. The competition between frustration and the Kondo effect was recently investigated by magnetic field tuning~\cite{Lukas,Akito}. In the isostructural Kondo lattice YbAgGe successive magnetic transitions were found below 1~K~\cite{umeo-jpsj04,budko-prb04}. Detailed measurements in magnetic field revealed distinct ordered phases A, B, C and D and a first-order metamagnetic transition between C and D~\cite{Schmiedeshoff-prb11}. However, its critical temperature is very close to zero, leading to quantum bi-critical scaling over a wide region in phase space, which has been related to the effect of strong geometrical frustration~\cite{tokiwa-prl13_2}. 

CeRhSn is another Kondo lattice crystallizing in the hexagonal ZrNiAl structure. In contrast to CePdAl and YbAgGe it remains paramagnetic upon cooling down to at least 50~mK~\cite{Kim-prb03,Schenck-jpsj04}. Divergences of the specific heat coefficient, Gr\"uneisen ratio and magnetic Gr\"uneisen parameter indicate that the system is in close vicinity to a zero-field QCP~\cite{Tokiwa-SA15}. Particularly striking is the anisotropy of the linear thermal expansion at low temperatures. The expansion coefficient $\alpha/T$ diverges only within the $ab$-plane while along the $c$-axis, it displays Fermi liquid behavior. The data indicate an anisotropic response of the entropy to stress (in the limit of zero stress). In particular the quantum critical entropy contribution has no $c$-axis stress dependence~\cite{Tokiwa-SA15}. Such behavior is in sharp contrast to quantum criticality in various other Kondo lattices where $\alpha/T$ diverges along all directions~\cite{KuchlerR:DivtGr,KuchlerR:Grurdq,Kuchler-PRL06,Gegenwart-Grueneisen-review,Grube2017}. The insensitivity of quantum critical entropy to initial stress along the $c$ direction, which leaves geometrical frustration unchanged, indicates a frustration-induced QCP in CeRhSn ~\cite{Tokiwa-SA15}. This material is thus well suited to investigate whether uniaxial stress can act as novel tuning parameter for quantum criticality in frustrated magnets. 

Smart devices with piezoelectric actuators, allowing in-situ tuning of strain, have recently been utilized for electrical resistivity and magnetic ac-susceptibility measurements on unconventional superconductors~\cite{Chu2012,Steppke2016}. However, thermal expansion and magnetostriction as most sensitive thermodynamic probes of phase transitions, depending on the uniaxial pressure derivatives of entropy and magnetization, respectively, are better suited to investigate frustrated Kondo lattices~\cite{Gegenwart-Grueneisen-review}. The data reported below provide evidence that uniaxial stress can reduce geometrical frustration and thereby opens a new possibility for tuning quantum critical states in frustrated magnets.

\begin{figure}[ht]
\includegraphics[width=\linewidth,keepaspectratio]{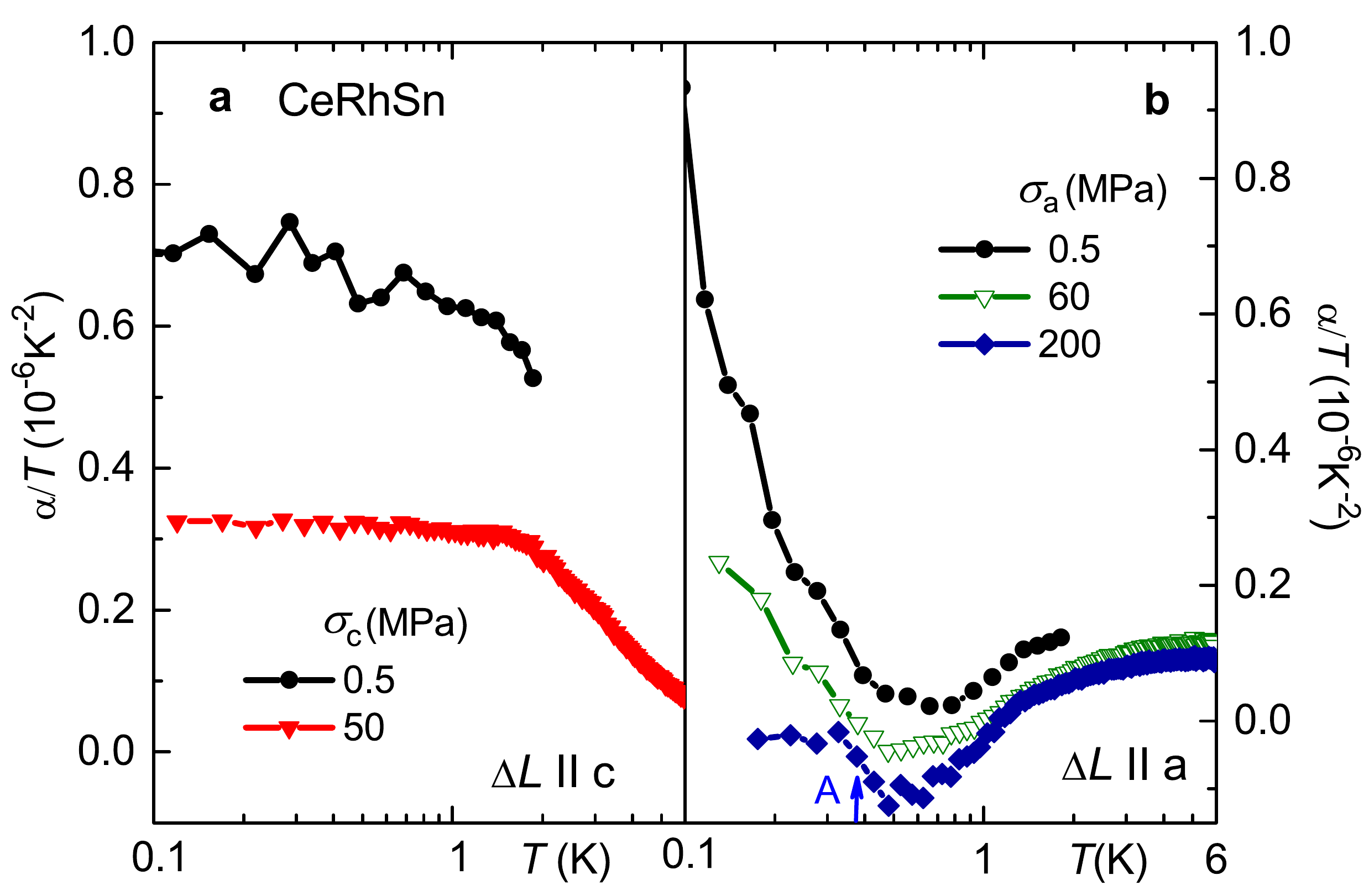}
\caption{Thermal expansion coefficient divided by temperature $\alpha/T$ measured along the $c$- (a) and $a$-axis (b) as a function of temperature at various different values of uniaxial stress applied parallel to the measurement direction. The data at 0.5~MPa have been taken from~\cite{Tokiwa-SA15}. The blue arrows indicate the $A$ cross-over and $A$-$B$ transition for 200~MPa, cf. Fig. 4.}
\end{figure}

Our measurements were conducted with a specially designed uniaxial stress capacitive dilatometer which allows to detect length changes as small as $0.02$~\AA~for a sample with length of order 1~mm~\cite{Kuchler-RSI}. The sample is clamped between a movable and a fixed plate by springs that exert a strong force of 55~N. The applied stress is calculated by dividing the force by the sample cross section. Since we used a single crystal of rectangular shape with the same cross section at both ends, a uniform stress along the sample is guaranteed which is an important advantage compared to piezo-strain devices. All measurements were performed on a well characterized single crystal studied previously~\cite{Tokiwa-SA15}. The applied stress was enhanced by subsequently reducing its cross section. The sample for the highest stress 200\,MPa has a cross-section of 0.5 x 0.6\,mm$^2$ and a length of 3.5\,mm. Magnetic fields $\bm{B}$ and uniaxial stress $\bm{\sigma}$ were always applied parallel to the measurement direction.
%The CeRhSn single crystals were grown by the Czochralsky method from the melt of stoichiometric amounts of the constitute elements in an rf induction furnace. The crystals were wrapped with tantalum foil, sealed in an evacuated quartz tube, and annealed at 900$^{\circ}$C for three weeks. Within the 1\,\% accuracy of electron microprobe analysis on our crystals no deviation from the nominal stoichiometry was found.

Previous low-temperature thermal expansion measurements on CeRhSn with the standard dilatometer, i.e., under very low uniaxial stress of 0.5~MPa, revealed a pronounced anisotropy~\cite{Tokiwa-SA15}. While the $c$-axis expansion coefficient $\alpha_c/T$ saturates at low temperatures, indicative of dominating Fermi liquid behavior, the $a$-axis coefficient $\alpha_a/T$ diverges upon cooling (cf. the black circles in Fig.~1a and b). This has been associated with a QCP related to geometrical frustration, which is sensitive only to in-plane but not to $c$-axis stress. We now turn to uniaxial-pressure measurements. As shown in Fig.~1a, a 
%In Figs. 1a and b, zero-field linear thermal expansion measurements along the $c$- and $a$-axis, respectively, are shown for various values of stress applied along the measured directions. The previous $c$-axis expansion data at low stress have shown saturation in $\alpha/T$ below about 1~K~\cite{Tokiwa-SA15}. This indicates the insensitivity of the quantum critical entropy to uniaxial stress along the $c$ direction, which leaves geometrical frustration unchanged.
stress of 50~MPa along $c$ reduces $\alpha/T$ by a factor of two but does not change its temperature dependence, in accordance with the presumption of unaffected geometrical frustration. The positive sign of the $c$-axis thermal expansion indicates, via the Gr\"{u}neisen relation~\cite{Gegenwart-Grueneisen-review}, that uniaxial pressure along $c$ enhances the dominating energy scale of the system. Indeed for Ce-based Kondo lattices, an increase of the Kondo scale with (uniaxial) pressure is expected. Along the $a$-axis, the previously measured positive $\alpha_a(T)>0$ at very small uniaxial pressure of 0.5~MPa indicates an enhancement of Kondo temperature also for $a$-axis stress. Without a release of geometrical frustration by distortion of the quasi-Kagome lattice~\cite{Tokiwa-SA15}, $a$-axis stress therefore cannot induce magnetic order in CeRhSn.

\begin{figure}[htb]
\includegraphics[width=\linewidth,keepaspectratio]{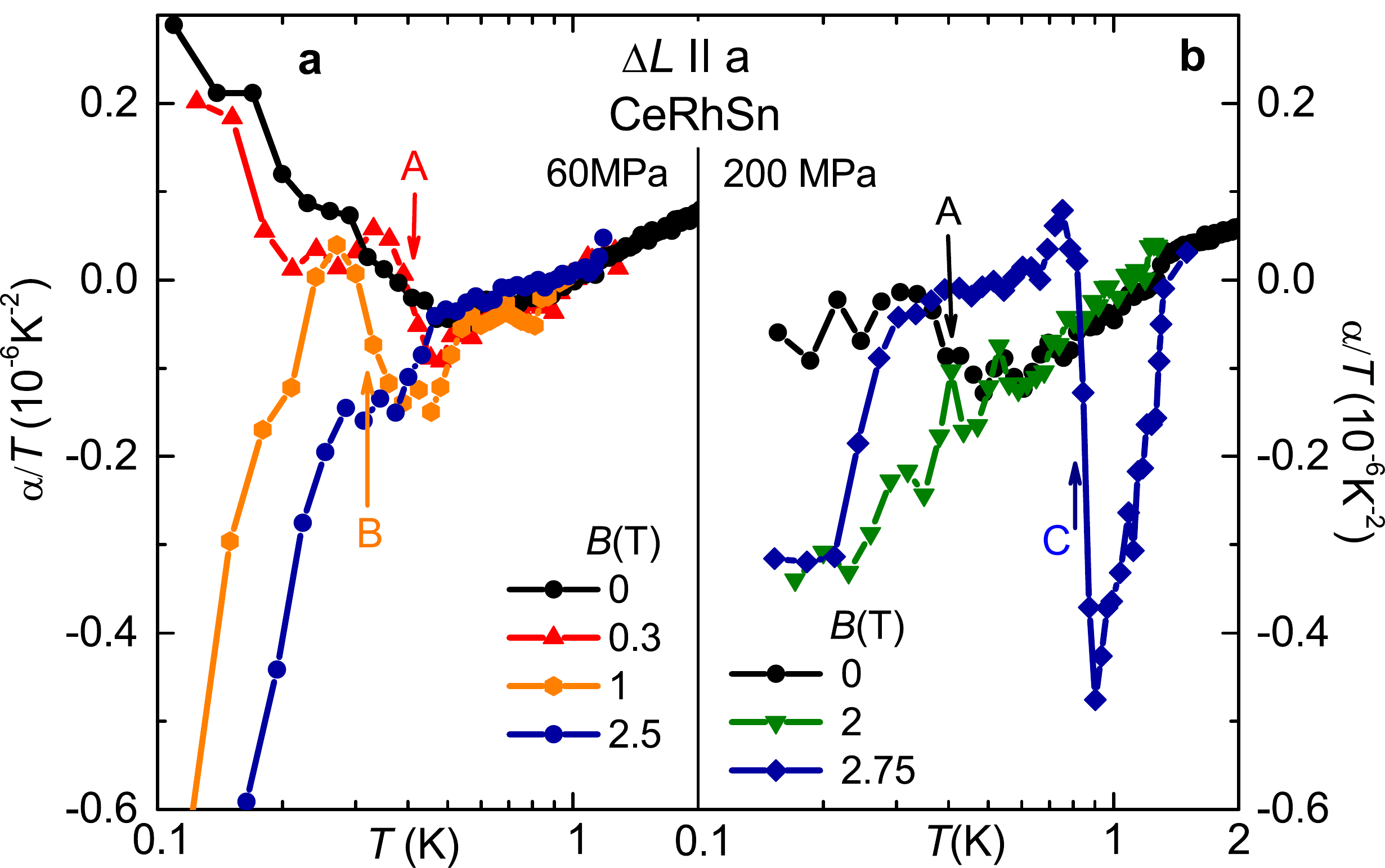}
\caption{Linear thermal expansion coefficient along the $a$-axis divided by temperature under stress $\sigma_a$ along the $a$-axis of 60 (a) and 200\,MPa (b) at various different magnetic fields, applied along the $a$-axis. Arrows indicate phase transitions (cf. also Fig.~4).}
\end{figure}

In contrast to $c$-axis uniaxial stress, which leaves geometrical frustration unchanged, $a$-axis stress, as shown in Fig.~1b, leads to a strong change of the temperature dependence of thermal expansion, in particular (i) a sign change of $\alpha(T)$ near 1~K indicating entropy accumulation (since $\alpha_a\sim (-dS/d\sigma_a)_T$, $S$: entropy), (ii) a suppression of the low-temperature divergence in $\alpha/T$, and (iii) a step-like change of the expansion coefficient at $T_A=0.38$~K (cf. the blue arrow), indicating a second-order phase transition. Generically, a zero-crossing of thermal expansion indicates a local maximum of entropy and arises at a magnetic phase transition near the QCP~\cite{GarstM:SigctG}. Therefore, these observations demonstrate that upon reducing geometrical frustration by in-plane stress CeRhSn transforms from the quantum critical into a long-range ordered state at low temperatures. In contrast to e.g. CeIn$_{3-x}$Sn$_x$ with $x=0.55$~\cite{Kuchler-PRL06}, the temperature where thermal expansion changes sign does not coincide with the second-order phase transition but is significantly larger. Furthermore, already the ``ambient'' pressure thermal expansion (at very small stress $\sigma_a=0.5$~MPa) shows non-monotonic behavior and decreases upon cooling from 2~K to 0.8~K. This indicates a negative contribution to thermal expansion that is strengthened by an increase of $a$-axis stress. Since, as detailed above, negative thermal expansion for Ce-based Kondo lattices cannot be explained by the pressure dependence of the Kondo effect, it very likely is related to magnetic correlations. We therefore associate the sign change in $\alpha_a(T)$ with short-ranged correlations or order preceding the magnetic order at $T_A$. Note, that isostructural YbAgGe also displays a broad crossover at $\sim 1$~K followed by a sharp magnetic phase transition near 0.5~K~\cite{Tokiwa-PRB06}.

We now turn to in-plane stress data at finite magnetic field (applied along the $a$-direction). As shown in Fig.~2, several step-like changes of $\alpha(T)$, indicative of second-order phase transitions, are resolved. They correspond to the transitions to different phases A, B and C which will later be discussed in a phase diagram. At a moderate stress of 60~MPa the data at 0 and 0.3~T display a positive increase at lowest temperatures, while for 1 and 2.5~T a negative divergence upon cooling to very low temperatures is found. This suggests distinct phases A and B. As will be shown below, they are separated by a metamagnetic transition. At 200~MPa this feature is absent and sharp negative peaks for 2.75 and 3~T indicate another field-induced phase transition labelled C. We do not have enough data at various different applied fields, to determine complete temperature-field phase diagrams at different applied strains. This is also related to the fact, that we can only change the applied strain by changing the cross-section of our crystal. In analogy to YbAgGe~\cite{Schmiedeshoff-prb11}, CeRhSn under $a$-axis stress may also show complicated phase diagrams with various magnetic phases. While more detailed determination of these phases requires much future work, the data at hand prove that quantum critical CeRhSn is tuned to an unidentified ordered state by $a$-axis stress. Since this cannot be explained by the strain effect on the Kondo interaction (see above), it has to be ascribed to a reduction of geometrical frustration.

\begin{figure}[htb]
\includegraphics[width=\linewidth,keepaspectratio]{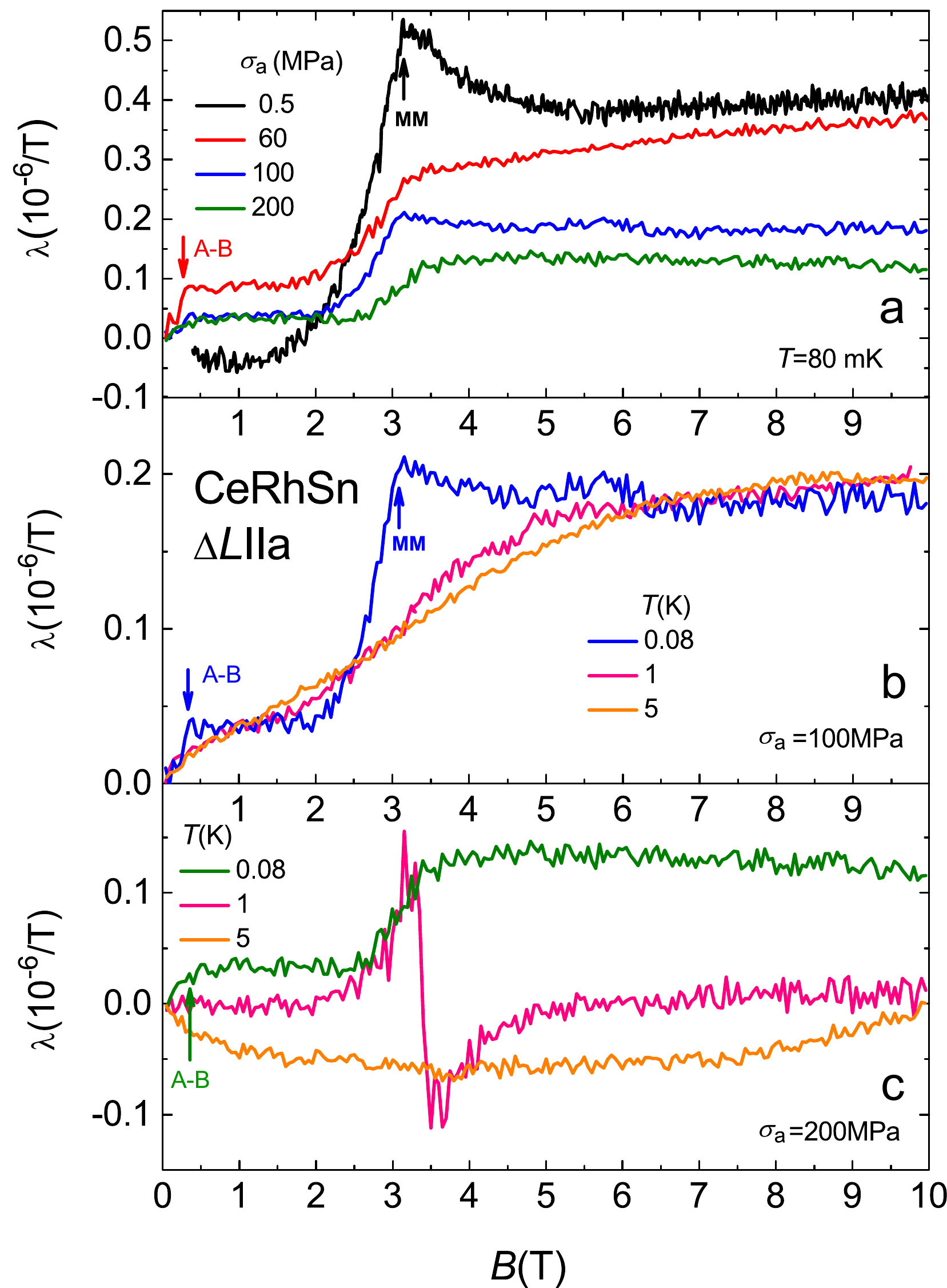}
\caption{Isothermal $a$-axis magnetostruction $\lambda=d(L_a/L_0)/dB$ ($L_a$: length of the sample along the $a$-direction) at 80~mK for different $a$-axis uniaxial stress (a) and at various different temperatures for 100 (b) and 200~MPa (c). The arrows indicate the transitions from phases A to B as well as the metamagnetic (MM) crossovers (cf. also Fig.~ 4).}
\end{figure}

Further information is obtained from magnetostriction measurements along the $a$-direction. Figure~3a shows the evolution of $\lambda$ at 80\,mK with increasing uniaxial stress $\sigma_a$. At low stress the peak around 3\,T indicates a MM cross-over, reported before~\cite{Tokiwa-SA15}. Because a metamagnetic transition of itinerant moments occurs along the axis of larger magnetization, while for CeRhSn it is opposite, this signature was instead interpreted as a spin-flop of local moments in a spin liquid state~\cite{Tokiwa-SA15}.
%We note that $\lambda_a(B)$ near 3.5~T looks very similar to magnetic susceptibility $\chi(B)$ around the metamagnetic transition at 4.5~T in YbAgGe~\cite{Tokiwa-PRB06}. For the latter material, a low-lying critical end point of the first-order metamagnetic transition enables pronounced quantum-bi-critical fluctuations~\cite{tokiwa-prl13_2}. Near metamagnetic quantum criticality, indeed a proportionality of $\lambda(B)$ to $\chi(B)$ is expected from general scaling arguments~\cite{Weickert-prb10}. Thus, our data under uniaxial in-plane stress indicate a striking similarity to YbAgGe at ambient pressure.
Uniaxial stress along the $a$ direction reduces the maximum at the MM transition (cf. Fig.~3b). The shape of the anomaly in $\lambda(B)$ at 3~T at 100 or 200~MPa more looks like a broadened step, which would be characteristic for a (smeared) second-order phase transition. A sharper step is found at very small fields of 0.3~T and is associated to the transition from phase A to B in the phase diagram (cf. Fig. 4). Furthermore, for 200~MPa stress complicated behavior is found in the region of the MM transition at elevated temperatures. As shown in Fig.~3c a sharp downwards step around 3.3\,T to negative values in $\lambda(B)$ arises at 1\,K. This feature is found to coincide with the C-phase transition in thermal expansion (cf. the phase diagram shown in Fig.~4).

\begin{figure}[htb]
\includegraphics[width=\linewidth,keepaspectratio]{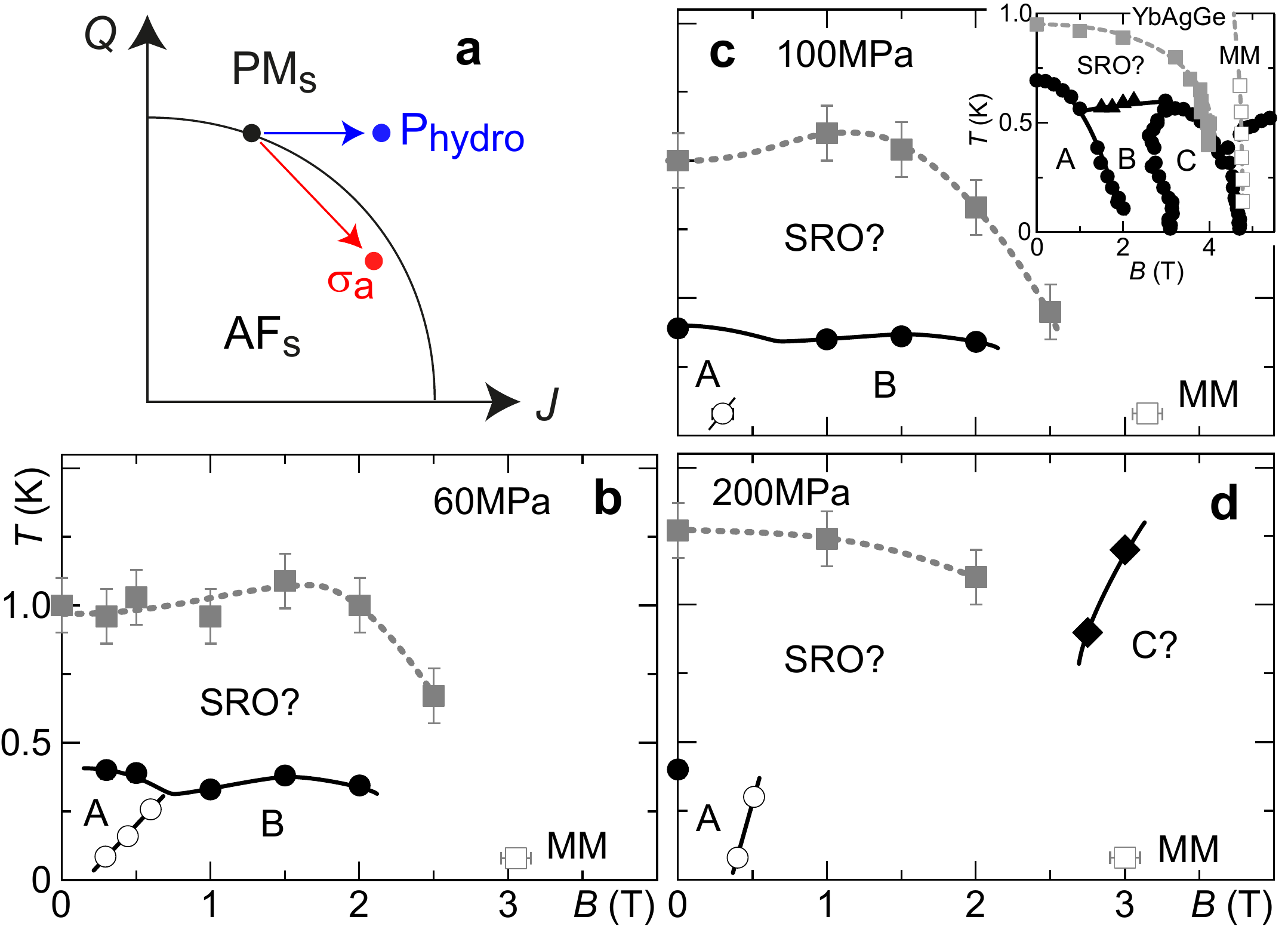}
\caption{(a) Schematic frustration ($Q$) vs. Kondo coupling ($J$) phase diagram at zero temperature with a quantum critical line, separating antiferromagnetic and paramagnetic ground states with small Fermi surface (labelled AF$_s$ and PM$_s$, respectively). For the complete ``global phase diagram'', see 
~\cite{Si-PhysicaB06}. The black dot denotes a possible location of CeRhSn at ambient conditions according to~\cite{Tokiwa-SA15}. The blue and red arrows indicate the increase of $J$ under hydrostatic pressure and the possible change of $Q$ and $J$ by $a$-axis stress in CeRhSn, respectively. (b),(c) and (d) display temperature versus magnetic field diagrams of CeRhSn under different values of $a$-axis stress for $B\parallel a$. Solid and open symbols are obtained from thermal expansion and magnetostriction, respectively, indicating phase transitions to presumed A, B and C phases. The grey dotted solid squares indicate the zero crossing of thermal expansion, possibly associated with short range order (SRO). MM denotes metamagnetic cross-over. The lines serve as a guides to the eye.
The inset in (b) shows phase diagram of isostructural YbAgGe with magnetic phases A, B and C and MM crossover~\cite{Schmiedeshoff-prb11}. The grey solid squares represent cross-over anomalies observed by electrical resistivity and specific heat measurements.}
\end{figure}

In panels (b) to (d) of Figure 4 the previously discussed crossover-signatures and phase transitions anomalies for three different values of the applied $a$-axis stress in CeRhSn are summarized. While there are similarities in all phase diagrams, details of the nature of these states, their phase boundaries as well as their stress dependence remain to be determined by future experiments. The complexity of the magnetic phase diagram of CeRhSn under in-plane uniaxial stress shares similarities to that of isostructural YbAgGe under ambient conditions, shown in the inset of Fig.~4c~\cite{Schmiedeshoff-prb11,tokiwa-prl13_2}. Upon cooling, thermal expansion displays a sign change at 1.2~K, indicating entropy accumulation, but no second-order phase transition. For YbAgGe, near 1~K a crossover to a possibly short-range ordered state has been observed~\cite{Tokiwa-PRB06}. Furthermore, induced by stress a series of distinct phases below 0.5~K is found, as in YbAgGe at ambient conditions. The complicated phase diagram with several competing ordered states in YbAgGe has been associated to magnetic frustration~\cite{umeo-jpsj04}. CeRhSn features a zero-field QCP at ambient conditions, driven by geometrical frustration~\cite{Tokiwa-SA15}. The change from a quantum critical into an ordered ground state under in-plane stress, despite the enhancement of the Kondo coupling, could only be explained by the influence of a release of magnetic frustration as sketched in Fig. 4(a). Further experiments on CeRhSn under in-plane stress would be interesting, in particular measurements of the magnetic susceptibility and heat capacity to quantify frustration, as recently performed in CePdAl~\cite{Lukas}, and neutron scattering to determine the nature of the different ordered states. In addition, studies of the Fermi surface would be highly interesting to determine the nature of the 4f-electrons in CeRhSn (i.e. localized or itinerant) and its variation across the phase diagram.

We have performed thermal expansion and magnetostriction experiments on the frustrated Kondo lattice CeRhSn under uniaxial stress. For stress applied within the Kagome plane clear phase transition anomalies are observed at 0.38~K, suggesting the development of magnetic order due to a release of frustration. Measurements under in-plane stress and magnetic fields reveal a complex phase diagram with several magnetic phase boundaries and crossovers that shares many similarities to that of isostructural YbAgGe at ambient pressure. Since the Kondo coupling increases with stress along either direction, the inducement of order out of a quantum critical state could only be related to the reduction of geometrical frustration. Thus, uniaxial stress has been successfully utilized as new control parameter for tuning a quantum critical geometrically frustrated material into a magnetically ordered ground state. This is in accordance with predicted ``global phase diagram'' in which geometrical frustration and Kondo coupling act as two independent tuning parameters~\cite{Si-PhysicaB06}. While for CeRhSn, our experiments support that the zero-field QCP at ambient conditions is driven by strong frustration~\cite{Tokiwa-SA15}, more generally, our study opens a new way for investigating geometrically frustrated matter. It could be expected, that in several other quantum spin liquid candidate materials novel hidden quantum phases can be discovered by a release of frustration with uniaxial stress.

%We are grateful to M. Garst and Q. Si for helpful discussions.
This work has been partly supported by the German Science Foundation through the grants KU 3287/1-1 and GE 1640/8-1 (``Tuning frustration in spin liquids by uniaxial pressure'') as well as by JSPS KAKENHI Grants No. JP26400363 and No. JP16H1076.

\bibliography{hf}

\end{document}